\documentclass[runningheads]{svmult}

\usepackage{makeidx}   % allows index generation
\usepackage{graphicx}  % standard LaTeX graphics tool
\usepackage{subeqnar}  % subnumbers individual equations
\usepackage{multicol}  % used for the two-column index
\usepackage{physprbb}  % modified textarea for proceedings,
\makeindex             % used for the subject index
\begin{document}
\title*{Astro Particle Physics from Space}
%
% allows explicit linebreak for the table of content
%
%
\author{Roberto Battiston\inst{1}}
%
%\authorrunning{Roberto Battiston}
% if there are more than two authors,
% please abbreviate author list for running head
%
%
\institute{Dipartimento di Fisica and Sezione INFN, I-06121
Perugia Italia}
%\institute{Princeton University, Princeton NJ 08544, USA}

\maketitle              % typesets the title of the contribution

\begin{abstract}
We review  how some open issues on Astro Particle physics can be
studied by space born  experiments  in a complementary way to what
is being done at underground and accelerators facilities
\end{abstract}

\section{ At the beginning it was Astro Particle}
This figure shows the conference photo taken at the 1939 Chicago
Symposium on Cosmic Rays. Among the partecipants we find a quite
exceptional group of physicists: Kohloster, Bethe, Shapiro,
Compton, Teller, Eckart, Gouldsmit, Anderson, Oppenheimer,  Hess,
Wilson, Rossi, Auger, Heisenberg, Weehler and
\begin{figure}
\begin{center}
\includegraphics[width=1.0\textwidth]{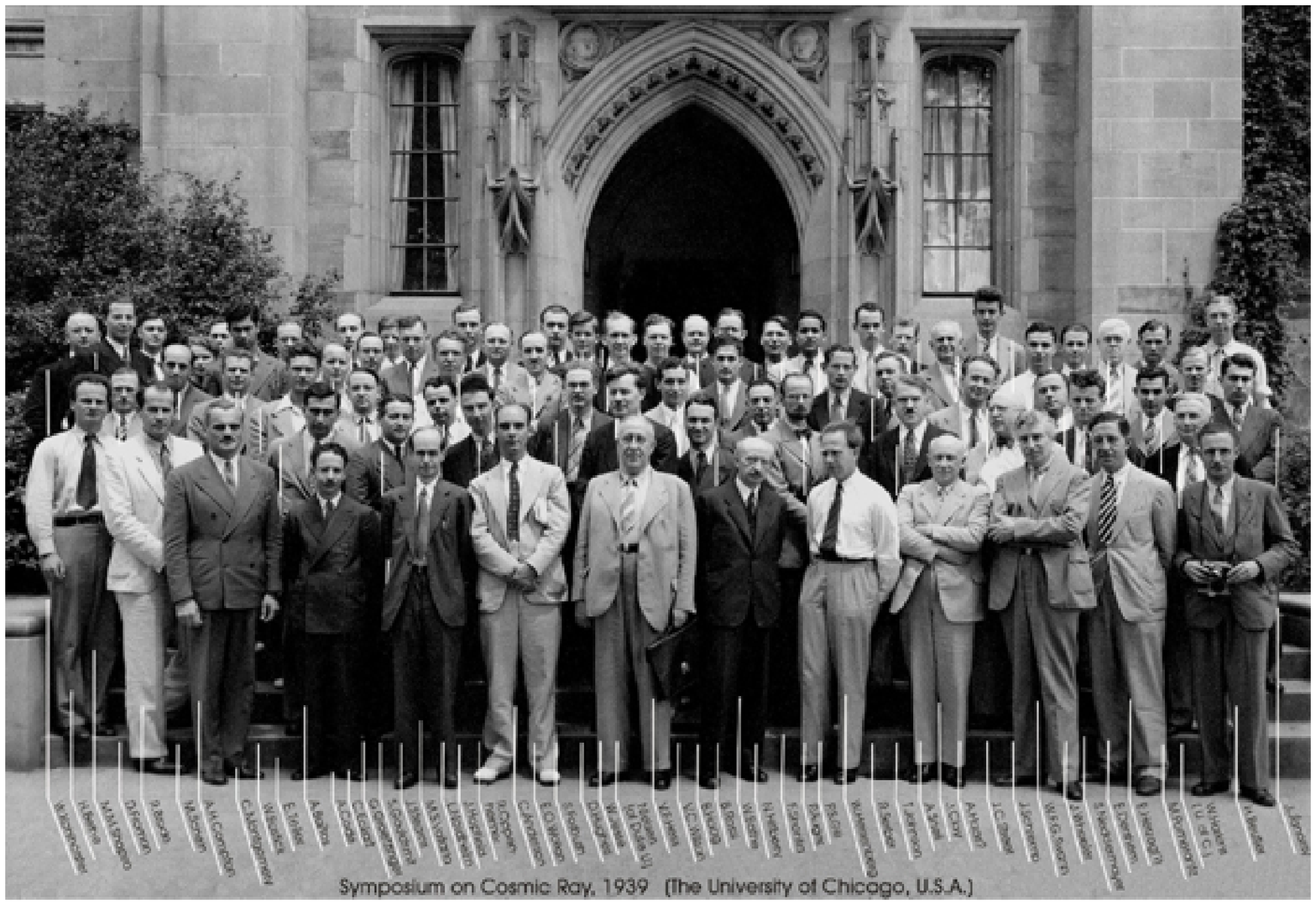}
\end{center}
%\caption[]{Conference photo from the 1939 Chicago Cosmic Rays
%Symposium}
\label{chicago}
\end{figure}

\footnote[1]{Invited Talk at the  ESO-CERN-ESA Symposium on
Astronomy, Cosmology and Fundamental Physics, March 4-7  2002,
Garching, Germany}

many other which are the among the fathers of the modern physics,
based on Quantum Mechanics, Elementary Particles and Fundamental
Forces. Why Cosmic Rays, discovered by Hess nearly 30 years before
were, still in 1939, such an interesting topics for these
distinguished scientists?

The answer lies in Table 1. In the years preceding 1937 both the
first antiparticle (the positron) and the first unstable
elementary particles $(\mu^{\pm})$ were discovered in Cosmic Rays.
Many more particles were to be discovered during the following
years analyzing the Cosmic Radiation, making Cosmic Rays symposia
very exciting until at least 1953, when experiments at
accelerators  started  to systematically discover new elementary
particles while Cosmic Rays experiments suddenly stopped finding
them.

\begin{table}
\caption{Discovery of elementary particles}
\begin{center}
\renewcommand{\arraystretch}{1.4}
\setlength\tabcolsep{5pt}
\begin{tabular}{llllll}
\hline\noalign{\smallskip}
Particle &  Year  & Discoverer (Nobel Prize) & Method \\
\noalign{\smallskip}
\hline
\noalign{\smallskip}
 $e^{-}$ & 1897 & Thomson (1906)& Discharges in gases \\
 $p$ & 1919 & Rutherford & Natural radioactivity \\
$n$ & 1932 & Chadwik (1935) & Natural radioactivity \\
$e^{+}$ & 1933 & Anderson (1936) & Cosmic Rays \\
$\mu^{\pm}$ & 1937 & Neddermeyer, Anderson  & Cosmic Rays \\
$\pi^{\pm}$ & 1947 & Powell (1950) , Occhialini& Cosmic Rays \\
$K^{\pm}$ & 1949 & Powell (1950) & Cosmic Rays \\
$\pi^{0}$ & 1949 & Bjorklund & Accelerator \\
$K^{0}$ & 1951 & Armenteros & Cosmic Rays  \\
$\Lambda^{0}$ & 1951 & Armenteros & Cosmic Rays  \\
$\Delta$ & 1932 & Anderson & Cosmic Rays  \\
$\Xi^{-}$ & 1932 & Armenteros & Cosmic Rays  \\
$\Sigma^{\pm}$ & 1953 & Bonetti  & Cosmic Rays \\
$p^{-}$ & 1955 & Chamberlain, Segre' (1959) & Accelerators \\
anything else & 1955 $\Longrightarrow$ today & various groups & Accelerators \\
$ m_\nu \neq 0$ & 2000 & KAMIOKANDE  & Cosmic rays \\
\hline
\end{tabular}
\end{center}
\label{Tab1a}
\end{table}

If Cosmic Rays have been instrumental  to give birth to  particle
physics during the first half of the past century, starting from
the fifties, however,  accelerators have been the tools for the
experimental  triumph of the Standard Model of Particle Physics,
including   the discovery of the electro-weak bosons at CERN or of
the  heavy sixth quark at Fermilab.

During the last ten years, however, the rate of discoveries  at
accelerators seems significantly reduced, possibly because of the
limited energy scale which can be tested at existing or future
facilities. A growing number of physicists is then turning  again
to CR with  new experimental techniques aiming to extend  by
orders of magnitude the sensitivities reached by past experiments.

In particular a  number of  space born experiments have  been
proposed to measure, with unrivalled  accuracy, the composition of
primary high energy CR, searching  for new phenomena not
accessible to present accelerators.

In this paper we  will review these experiments and their physics
potential.
\begin{figure}
\begin{center}
\includegraphics[width=.6\textwidth]{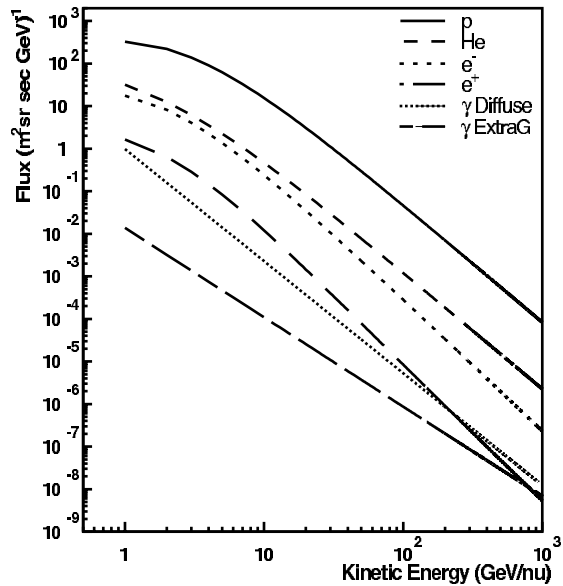}
\end{center}
\caption[]{High Energy Cosmic Rays composition \cite{choutko}}
\label{hecr}
\end{figure}
The paper is organized in two parts. In the first part I will
discuss the characteristics of the CR flux, the beam nature gives
us, reviewing the status of our knowledge of their energy spectrum
and composition. In the second part I will discuss some of the
space born experiments planned in the next future which will
contribute to the quest for  answers to various unsolved questions
in Astro Particle physics.

\section{Understanding Nature's Beam}

The Universe communicate with us by sending to Earth  a continuous
flow of radiation of different kind. Here we are  interested to
the high energy  part of the spectrum ($E > O(GeV)$). Cosmic Rays
traditionally refer to the charged component of high the energy
particles travelling through the galaxy (Figure \ref{hecr}):  in
order of abundance $p \sim 80 \% , ^{4}He \sim 15 \%,  e^{-}: O(1
\%), e^{+}: O(0.1 \%),
 \bar{p}:O(0.01 \%)$). In addition to  $p$ and $^{4}He$, there is a composite  hadronic  component
  including  all other nuclei
 and  long lived isotopes, totalling about a few $\%$ of the   total CR flux.

 There are however  two other forms of energetic radiation which are relevant for
  Astro Particle physics.
The first is the high energy part of the electromagnetic spectrum,
gamma rays above  $\sim 100\ MeV$. Their flux  is at the level of
$10^{-5}$ of the CR flux. Gamma rays have an important property
that charged CR do not have, namely  they travel along straight
lines, undisturbed by the magnetic field and  reproducing the
images of their sources.
%\vskip -1cm
  \begin{figure}
\begin{center}
\includegraphics[width=1.0\textwidth]{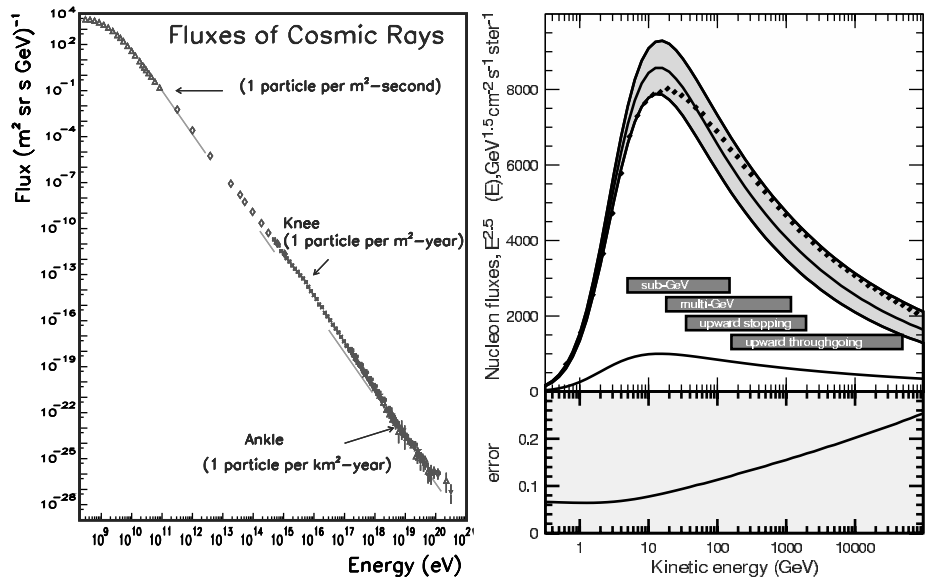}
\end{center}
%\vskip -0.5cm
\caption[]{Left: the flux of charged cosmic rays.
Right: measurement and  uncertainty  on the primary  CR spectrum
(proton and He).The accurate knowledge of this part of the CR
spectrum is important for the understanding of    the atmospheric
neutrino flux \cite{stanev}.} \label{stanev}
\end{figure}

  The second  component  are  the neutrinos, which   would also  allow  source imaging:
  at high energy the primary neutrino flux is much lower and steeper with  increasing energy
    than  for charged CR, with the exception  of
   the secondary  atmospheric component  induced by high energy CR hitting the atmosphere.
    Around $1 \ GeV$ the neutrino flux is of the order of  the  flux of high
    energy gamma rays, but because of their low cross section, $\nu 's$ are  much harder to
detect.

    If we are plan  to use this energetic radiation to search for new particles or new effects, we must know the
    properties of Nature's beam with the best possible accuracy.
    Which is our current level on the knowledge of the Cosmic Radiation  and what we can expect
    in the coming future?

    \section {The charged CR component}

    \subsection{Charged  hadrons}

    Over  the last 40 years the hadronic CR  component has been measured systematically, using balloons (mostly),
     space born (sometimes), and ground based detectors. The most complete information is obtained by particle
     spectrometers, experiments able to measure directly the CR composition  through the determination of the charge,
     the sign of the charge, the momentum and the velocity of each particle. Often, however, simpler
     apparatus were used, based on calorimeters, emulsion chambers or Cerenkov detectors: in these cases only
     partial information  were obtained.  Until recently the most sophisticated CR spectrometers were flown
      on balloons operating between 30 and 35 km  of height (BESS\cite{BESS}, MASS\cite{MASS}, CAPRICE\cite{CAPRICE},
      IMAX\cite{IMAX}) but in 1998 a large magnetic spectrometer, AMS\cite{AMSPR}, has been operated  in space, providing for the first time a very precise
measurement on the composition  of primary CR before their
entrance in  the atmosphere.

%\vskip -1cm
\begin{figure}
\begin{center}
\includegraphics[width=.7\textwidth]{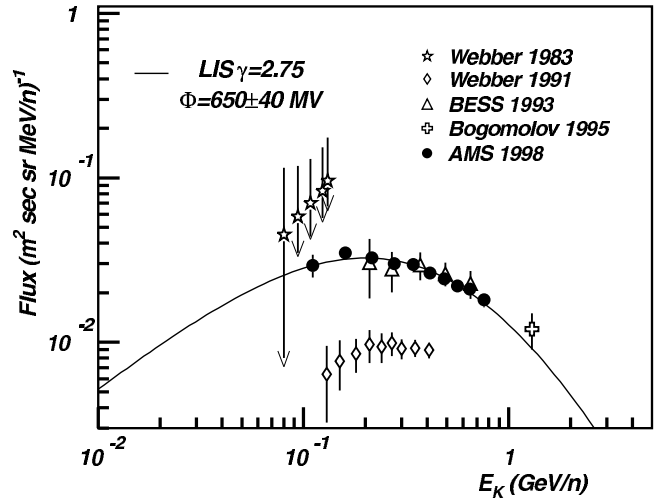}
\end{center}
%\vskip -1.5cm
 \caption[]{Cosmic deuteron
 measurements\cite{webber1,webber2,BESSde,bogolomov,lamanna}}
\label{deuteron}
\end{figure}

  The flux of the main  CR component,
     the  protons,  is known with $5-10 \%$ accuracy up to $\sim 200\ GeV$, and with $10-30\%$ accuracy up to
      $\sim 100\ TeV$ (Figure \ref{stanev}) \cite{stanev}.  Helium flux is known with $10 \%$ accuracy up to $\sim 10\ GeV$ but above this energy the
      measurements are rather poor. Light  $Z > 2 $ nuclei have been measured with about $5 \%$ accuracy only
       up to $\sim 35\ GeV$. It would be important to extend the energy range for precise measurements of
       hadrons up to about $10\  TeV$ because they are the source of the atmospheric neutrinos used for the determination
       measurement of the neutrino mass in underground experiments.

             For the study of CR composition, precise measurements of  some
                stable isotopes like Deuterium, Boron and Carbon or long-lived
             isotopes like $^9Be$ are particularly important to understand
              the propagation  and trapping mechanisms
             of CR in our Galaxy. Accurate measurements of
             $D$ are available only up a few $GV$ of rigidity (Figure \ref{deuteron})   while the
             knowledge of the  cosmologically important  ratio
             $^9Be/^{10}Be$ is very poorly known above about  hundred $MV$ of rigidity.

\subsection{Antiprotons}

\begin{figure}
\begin{center}
\includegraphics[width=.6\textwidth]{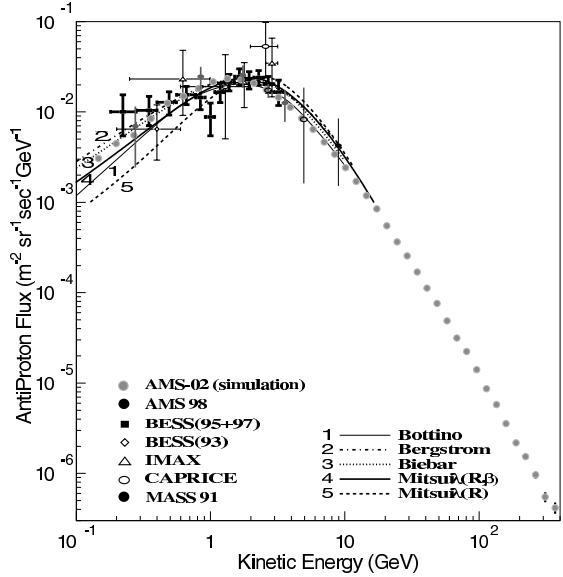}
\end{center}
%\vskip -2.2cm
 \caption[]{Compilation of antiproton measurements. A simulated measurement of the AMS-02 experiment  on the
 ISS is also reported. For all balloon data and models   see\cite{besspbar} and
 reference therein. For AMS data see\cite{AMSPR}}
 \label{antiproton}
\end{figure}

Antiprotons  are a rare but interesting  hadronic component of
high energy CR, because they could be produced by exotic
    sources like  antimatter dominated regions or  by the decay or annihilation of new particles.
    Their flux ratio to protons is  at the level of $O(10^{-4})$ at kinetic energies around $1\ GeV$.
     This rate is in agreement with the expectation
    that $\bar{p}$ are produced in high energy CR  interactions  with the interstellar medium. However in more than
    40 years of experiments with balloons, only a few thousands $\bar{p}$ have been measured, mostly with energies below
    $10\ GeV$ (Figure \ref{antiproton}). Statistical errors are then quite large, in particular below $1\ GeV$ and above
    $10\ GeV$: also systematic errors due to uncertainties in the modelling of the
propagation in the Interstellar Medium or of the solar  modulation
are still at the level of $20-30 \%$.

%\vspace{-1cm}
\subsection{The charged leptons}

\begin{figure}
\begin{center}
\includegraphics[width=1.\textwidth] {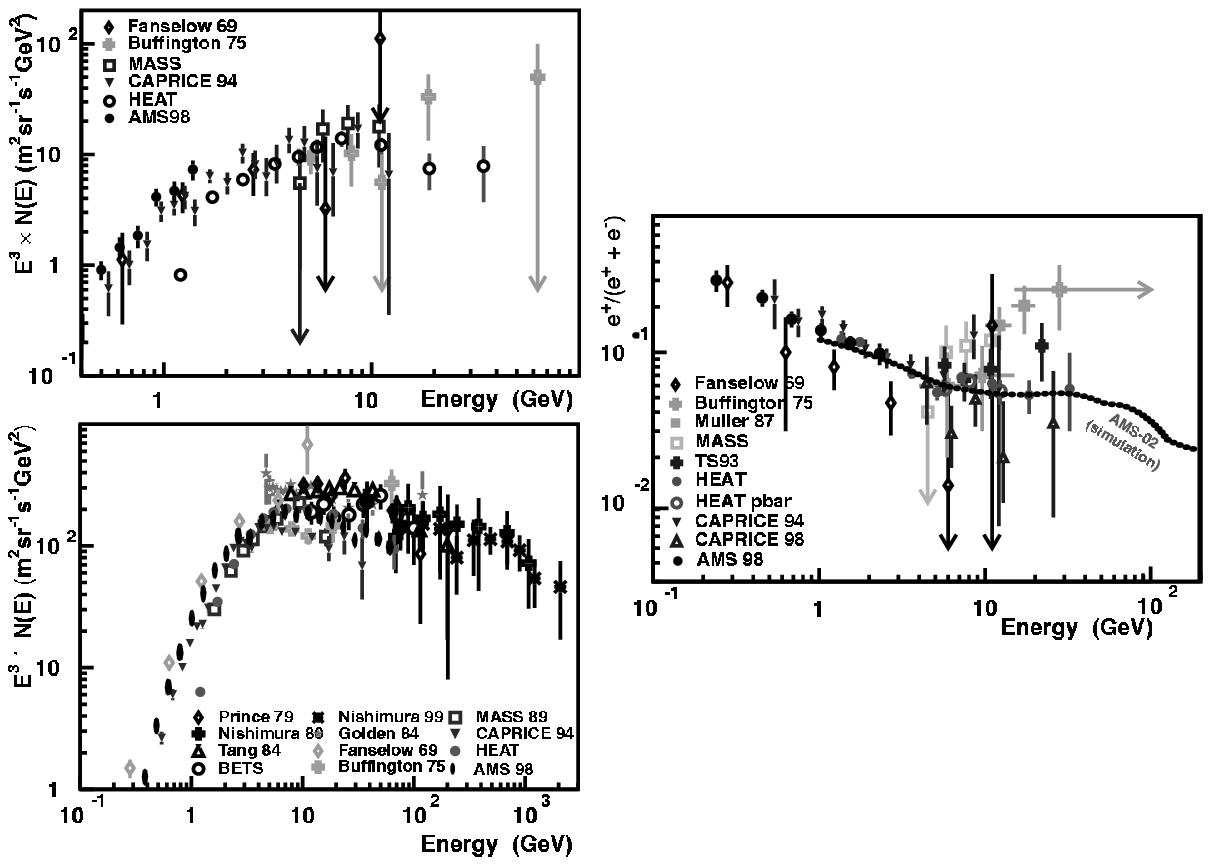}
\end{center}
\caption[]{Compilation of $e^+$ (top left), $e^-$  (bottom left)
and $e^+/(e^-+e^+)$ (right) data: labels corresponds to the
various references cited in \cite{bertucci}. A simulated long
exposure result from AMS-02 is also included, for the case
$m_\chi=275\ GeV/c^2$\cite{diehl}.  All experiments are balloon
borne spectrometers with the exception of AMS 98 and AMS-02 which
are space borne spectrometers} \label{electronpositron}
\end{figure}

Due to their lower flux ($\sim 0.5 \%$ for $e^-$ and $\sim 0.05
\%$ for $e^+$ around $1\ GeV$) and their steeply falling spectrum,
statistical uncertainties on the measurement of the spectra of the
two stable leptons are  larger;  the experimental situation gets
very confused above a few $GeV$ (Figure \ref{electronpositron}).
Precision measurement of electrons and positrons are important
since, being their fluxes quite low and with a strong charge
asymmetry, contribution form exotic sources like supersymmetric
particles annihilation  could distort the $e^+/e^-$ ratio at a
level which could detectable by an high precision
experiment\cite{diehl}.

\subsection{Future experiments}

During the next 5 years our knowledge of the flux and composition of CR will be greatly improved
       thanks to  two space borne spectrometers, PAMELA on a Russian rocket in 2003-2006\cite{picozza}
        and AMS-02 on the International Space Station  on 2006-2009\cite{battiston}. These two experiments will increase
        the statistical samples of  charged CR by a few to several  orders of magnitude. AMS-02, in particular,
         thanks to its  strong magnetic field and large geometrical factor,   will be able to extend  these measurements  well in to the $TeV$ region,
          covering a region which is interesting for various physics topics and which
today is very poorly known. For a discussion on the improvement
expected from AMS-02 on the hadronic CR component see
\cite{casaus}.
    Space born experiments operating for three years or more are much more accurate that balloon flights
    lasting only about one day. Long duration balloons experiments lasting for 20 days or more, however,
    could be competitive, as it is  shown in Table 2.
    \begin{table}
\caption{Future CR spectrometers}
\begin{center}
\renewcommand{\arraystretch}{1.4}
\setlength\tabcolsep{5pt}
\begin{tabular}{lllllll}
\hline\noalign{\smallskip}
  &  Aperture   &  Duration  & Altitude & Latitude & Launch & Area*Time  \\
  &  $(cm^2 sr)$  &  (days) & (km) & (degrees) & (year) & (AMS-01)  \\
\noalign{\smallskip} \hline \noalign{\smallskip}
  AMS-01 & 2300 & \ \ \ 10 & 320-390 & $< 51.7$ & 1998 & \ \ \ 1.0\\
PAMELA & \ \ \ 21  & 1000 & \ \ 690 & \ \ \ 70  & 2003 & \ \ \ 1.1 \\
BESS Polar & 3000 & \ \ \  20 & \ \ \ 36 &$> 70$ & 2004 & \ \ \ 2.6 \\
AMS-02 & 5000 & 1000  & 320-390 & $< 51.7$ & 2006  & 217.0 \\
\hline
\end{tabular}
\end{center}
\label{Tab2a}
\end{table}

While at high energy the AMS-02 will be the most performing
experiment,  the Polar BESS\cite{sanuki} long duration flights,
thanks to BESS large acceptance, will have a good sensitivity to
the lower energy part of the CR spectrum and will perform
    a precise measurement of low energy $\bar{p}$. On the other
    side the space  borne PAMELA spectrometer,  in spite of its small
    aperture, is a timely experiment  and  will have a chance to perform high
    statistic CR measurements  in the period between AMS-01 and AMS-02.

        At energies around  the  knee  ($\sim 10^{15} eV = 1\ PeV$) and above,  all measurements come nowadays from
         large area  ground  based   arrays.
        At these energies the  CR  elemental composition can be determined only by modelling
        the interaction of CR with the atmosphere
        and it is affected by large uncertainties. To solve  this problem a long duration,  large acceptance,
         space borne experiment, ACCESS, has been  considered\cite{ACCESS}  either for deployment on the International Space
         Station or as a free flyer.

Ground based large area arrays have  measured the CR spectrum (not
the composition) up to Extreme Energies ($10^{21} eV = 1\ ZeV$)
and they are at the moment the only experiments able to measure
these very rare events. At these energies the flux is so low that
eventually it will  becomes unpractical to extend the surface or
the time of exposure of ground based  experiments. The largest
experiment of this kind, Auger\cite{auger}, is being built right
now. The next step will, however, to look to the fluorescence
induced by the EECR showers from  space, as proposed by
EUSO\cite{scarsi}, an experiment planned on the International
Space Station and capable to collect 10-50 times more statistics
than Auger or by KLYPVE\cite{khrenov} planned on a Russian free
flyer.

\section{The neutral CR component}
\subsection{High energy gamma rays}
    High energy  gamma rays  are a rare ($O(10^{-5})$)  component of the cosmic radiation which is not traditionally
     included in a CR review paper.  However   high energy gamma rays are produced
    by the same sources producing high energy  CR and carry complementary information.
    They should then be considered when discussing Astro Particle physics, in particular  since their study
     could give  important contributions  to the  understanding the problem of the origin of dark matter.
    Their energy  spectrum could, in fact,  be influenced by exotic sources like  neutralino annihilations taking place
     at the center of the galaxy.
     Most of the high energy gamma rays data have been collected by the EGRET experiment on the CGRO satellite during the
     $90's$. Since the end  of the CGRO program ($1999-2000$) there are no experiments measuring high energy
     gamma rays in space.
     During the present  decade there will be three space borne  experiments which will be able to measure high energy
     gamma rays: AGILE\cite{tavani} (2003) a small scientific mission of the Italian Space Agency, AMS-02 on the International
     Space Station (2006)\cite{battistongamma}  and GLAST\cite{morselli} (2007). These experiments will be able to cover the region up to  $\sim 300\ GeV$
     competing with ground base Cerenkov detectors which meanwhile will try to lower their threshold
     below $\sim 50\ GeV$.

\begin{figure}
\begin{center}
\includegraphics[width=.7\textwidth]{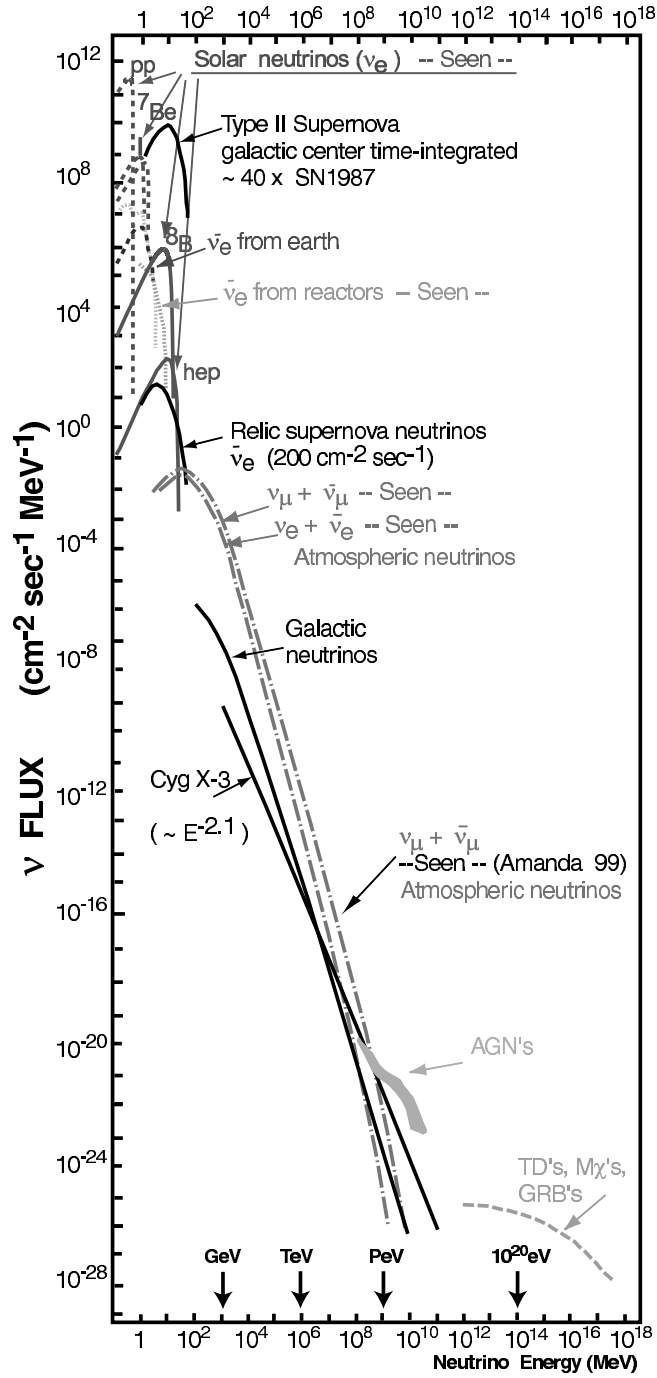}
\end{center}
\caption[]{The Cosmic neutrinos spectrum} \label{nuspect}
\end{figure}

\subsection{High energy neutrinos}

Neutrinos are a very important part of the Cosmic Radiation. Their
spectrum (Figure  \ref{nuspect}) extend over several orders of
magnitudes like for the charged  CR component, but is rich of
features coming from the various physical process at work.

\begin{figure}
\begin{center}
\includegraphics[width=.7\textwidth]{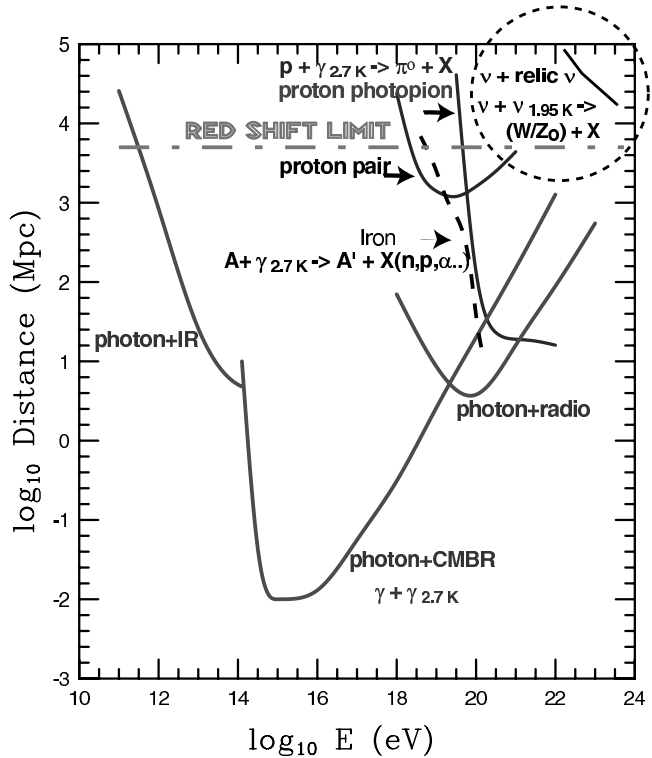}
\end{center}
\caption[]{Energy dependence of GZK cutoff for different CR
species\cite{letessier}} \label{ParticlePropagation}
\end{figure}

Neutrinos have a two great advantages and one disadvantage with
respect to  charged CR:

Advantages:
\begin{itemize}
\item 1) they travel on straight lines so  neutrino astronomy is a
possibility; \item 2) they have a small interaction  cross section
so they are basically unaffected by the GZK\cite{GZK} cutoff
 and can reach us from the edges of  the universe (see Figure \ref{ParticlePropagation}).
\end{itemize}

 Disadvantage:
 \begin{itemize}
\item
l) they have a small interaction  cross section   so their detection is problematic, requiring very large volumes
of matter.
\end{itemize}

Many of the neutrinos spectral features have been measured ($\nu$
from the sun, from  $SN1987$, from reactors, from the atmosphere),
some may be on the verge of being seen as   $\nu$'s of galactic
origin\cite{halzen}  while other are expected to exist but  are
still undetected (relic neutrinos, neutrinos from AGN's).  Some
could be produced by  exotic sources, like superheavy particles or
topological defects. The interest of neutrinos is their capability
to escape  the GZK cutoff up to extreme energies: they are the
only particle that can reach us from the edges of the universe,
carrying information about the status of space, time and matter,
before the recombination took place (Figure
\ref{ParticlePropagation}).

The main difficulty for neutrino detection is related to the mass
required for their detection. The only way to access energies of
the order of $10^{14}eV$ or above is to use the earth  surface
layers
 as targets,
like the atmosphere,  ice   or  water. In the case of the
atmosphere the fluorescence light emitted by the developing shower
could be detected from a space experiment able to monitor a
sufficiently large area of our planet.  This is the purpose or the
EUSO\cite{scarsi} experiment on the International Space Station,
expected to increase the sensitivity of  the Auger array by  more
than an order of magnitude by the end of the decade.

\section{Searching for new particles}

In this section  now briefly discuss the physics potential of the
accurate study of the cosmic radiation. An area of clear interest
for particle physics is obviously  the search of new states of
matter or for new particles.

   \begin{figure}
\begin{center}
\includegraphics[width=.6\textwidth]{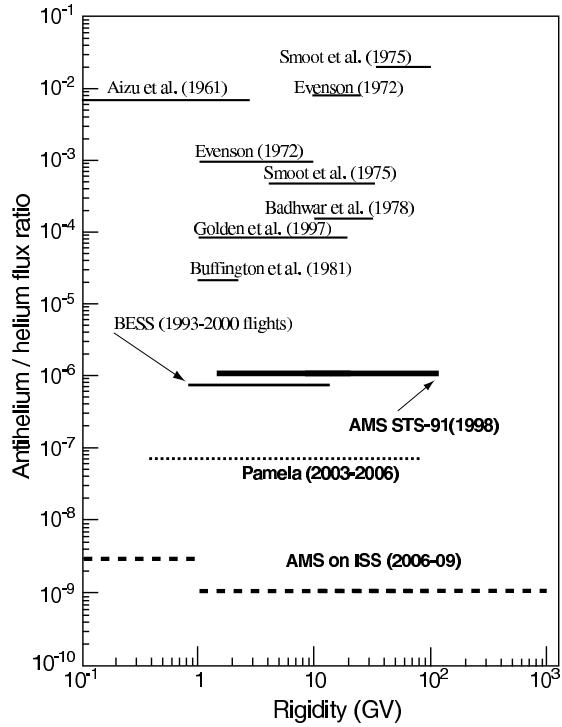}
\end{center}
\caption[]{Antimatter limits. For the references see
\cite{amsantimatter}} \label{antimatter}
\end{figure}

\subsection{Direct search for nuclear anti-matter}

The disappearence of  antimatter \cite{steigman,kolb,peebles}
 is one of the most  intriguing puzzles in our current understanding of the structure of the Universe.  Absence of
nuclear  antimatter from  the scale of our galaxy to  the scale of the local supercluster  is experimentally established at
 the level of $1$  part in $10^6$  by direct CR searches and indirect methods like the study of the energy spectrum
 of the diffuse gamma ray flux.   For a genuine antimatter signal one should look to nuclei heavier that $\bar{p}$
   since  secondary $\bar{p}$  can  be  easily produced at the level  $10^{-4}$  in high energy hadronic
 interaction of CR with the IM. This probability quickly vanishes with increasing atomic number.
For $\bar{D}$   the secondary production is at the level of $10^{-8}$ or less and
   already for $\bar{^4He}$ it  is  well below $10^{-12}$.

 \begin{figure} [ht]
\begin{center}
\includegraphics[width=.7\textwidth]{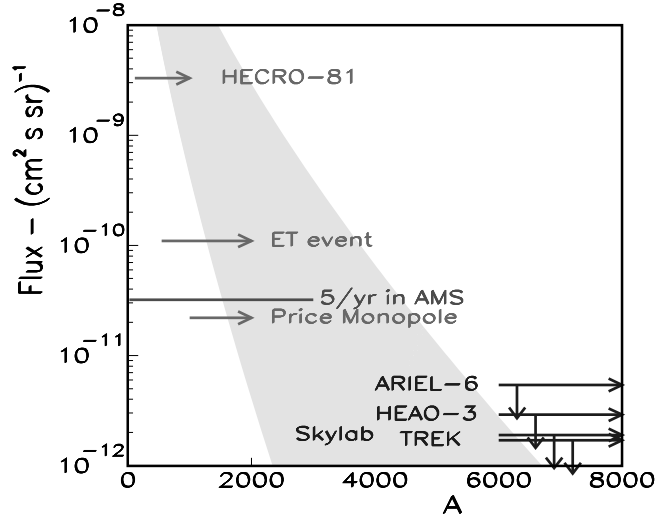}
\end{center}
\caption[]{Sensitivity to strangelets in AMS-02}
\label{AMSstrangelets}
\end{figure}

 This is why the unambiguous  observation of a couple of
   $\bar{^4He}$ at the level of one part in a billion or more would have profound implications on our understanding
   of baryogenesis. During the last  35 years experiments on balloons have pushed the limit on  the $\bar{^4He}$ at the
   level of less than about one part in a million (Figure \ref{antimatter}).
   Recently the AMS-01 spectrometer, during a 10 days
   precursor flight on the Shuttle, has reached the same level of  sensitivity. In the coming years PAMELA, first, and
   AMS-02, later, will  eventually  reach a sensitivity a thousand time better, reaching  rigidities of the order of a TV.

\subsection{Direct search for strange states of matter}
Strangelets\cite{witten} are a stable  state of nuclear matter
containing a large fraction of strange quarks. Such states could
have developed during the cosmological quark-hadron phase
transition $10^{-5} s$ after the Big Bang or in the high density
conditions of  compact supernova remnants, which might then be
strange stars composed of quark matter rather than neutron stars.
These stars could send to space fragments of stable strange
nuclear matter during catastrophic events like  collisions  with
other strange stars. These fragment would  have  variable mass but
a peculiar charge to mass ratio. For instance in the case of
color-flavor locked strangelets we obtain a charge to mass
relation of the type $Z= 0.3  A^{2/3}$. These particles would look
like  high mass, low mass/charge ratio cosmic rays which could be
easily identified in a space born magnetic spectrometer like
AMS-02. Figure \ref{AMSstrangelets} shows the sensitivity to
strangelets expected for  AMS-02  after three years on the
International Space Station\cite{madsen}.

\subsection{Indirect search for Dark Matter}

The presence at all scales in our universe of a non luminous
components of matter, Dark Matter (DM)\cite{ellis,turner},  is
possibly the most fascinating problem in Astro Particle physics. A
viable solution, if not the most viable solution\cite{masiero} to
this problem, is given by the Lightest Supersymmetric Particle
(LSP), the neutralino ($\chi$). Supersymmetry   links the existing
Standard Model particles to a set of new, heavier, super
 particles  through  $R$-parity  conservation, where $R$ is a combination  of the
 particle  spin, lepton and baryon numbers, $R= (-1)^{3B-L+2S}$.  The conservation of $R$-parity  requires that
 the LSP  is stable. LEP results suggests that the  LSP is heavy  ($m_\chi > 45\  GeV$)\cite{ellisneutralino} and
 then these particles can be a good DM candidate  in the Cold Dark Matter
(CDM) scenario.

% SUSY can be reduced to a 7  parameters theory: the $Higgsino$ and $gaugino$  mass parameters, $m$
% and $M_2$, the ratio  of the $Higgsino$ vacuum expectation values, $tan \beta$
% mass of the $CP$-odd Higgs, $m_A$, the scalar mass parameter, $m_0$ and the  trilinear soft SUSY
% breaking parameters  $A_b/m_0$, $A_t/m_0$.

Unfortunately  SUSY is a theory with many parameters  still poorly
constrained.
 The LSP (neutralino)  can be expressed as superposition of the neutral gauge ($g$ and $W$)
 and $Higgs$ boson superpartners ($H_{01}, H_{02}$):
\begin{equation}
             \chi = N_{11} B  +  N_{12} W_3  + N_{13} H_{01} + N_{14} H_{02}
\end{equation}
The parameters of this superposition define the $\chi$ properties, like the mass,
the annihilation cross section and branching ratios into detectable particles.
These parameters can also be related to the $\chi$  cosmological density, which
can be constrained in the interesting region $0.1 < \Omega _{DM} < 0.3$, in agreement
 with the  recent astrophysical results\cite{debernardis}.

 \begin{figure}
\begin{center}
\includegraphics[width=.9\textwidth]{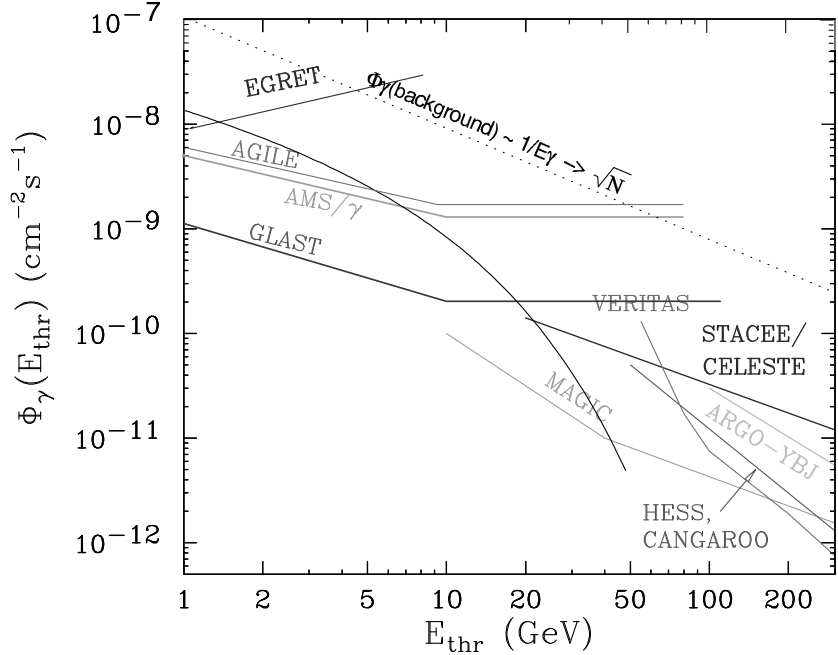}
\end{center}
\caption[]{Sensitivity  to SUSY dark matter for  various HE gamma
ray experiment, either ground based (bottom right) or space based
(top left). Vertical axis: gamma rays integral flux above a the
energy threshold $E_{thr}$. The dotted line represent the integral
flux of HE gamma rays from known sources, which represent a source
of background for  this measurement. The continuous line represent
the prediction for a  SUSY model, with $m_\chi = 120\ GeV$ and  a
boost factor $<J(0)>=5000$\cite{wilczek}.} \label{wilcekethr}
\end{figure}

 $\chi$ annihilation would take place in the most dense regions of our galaxy, e.g. its
 center or in other existing DM clumps. These regions  could be the source of prompt high energy CR without
 need of an acceleration mechanism.  In order to detect these prompt CR
 we study   rare  CR components  where the effects of the exotic $\chi$
 contributions would be detectable against backgrounds due to the primary
 spectrum.
  Rare CR components like high energy $\bar{p}$\cite{ellis}, $e^+$\cite{diehl}, $\bar{D}$\cite{donato}
   and $\gamma$\cite{bergstrom}  have been suggested as potential indirect  signatures for
 cold  $DM$.

\begin{figure}
\begin{center}
\includegraphics[width=0.7\textwidth]{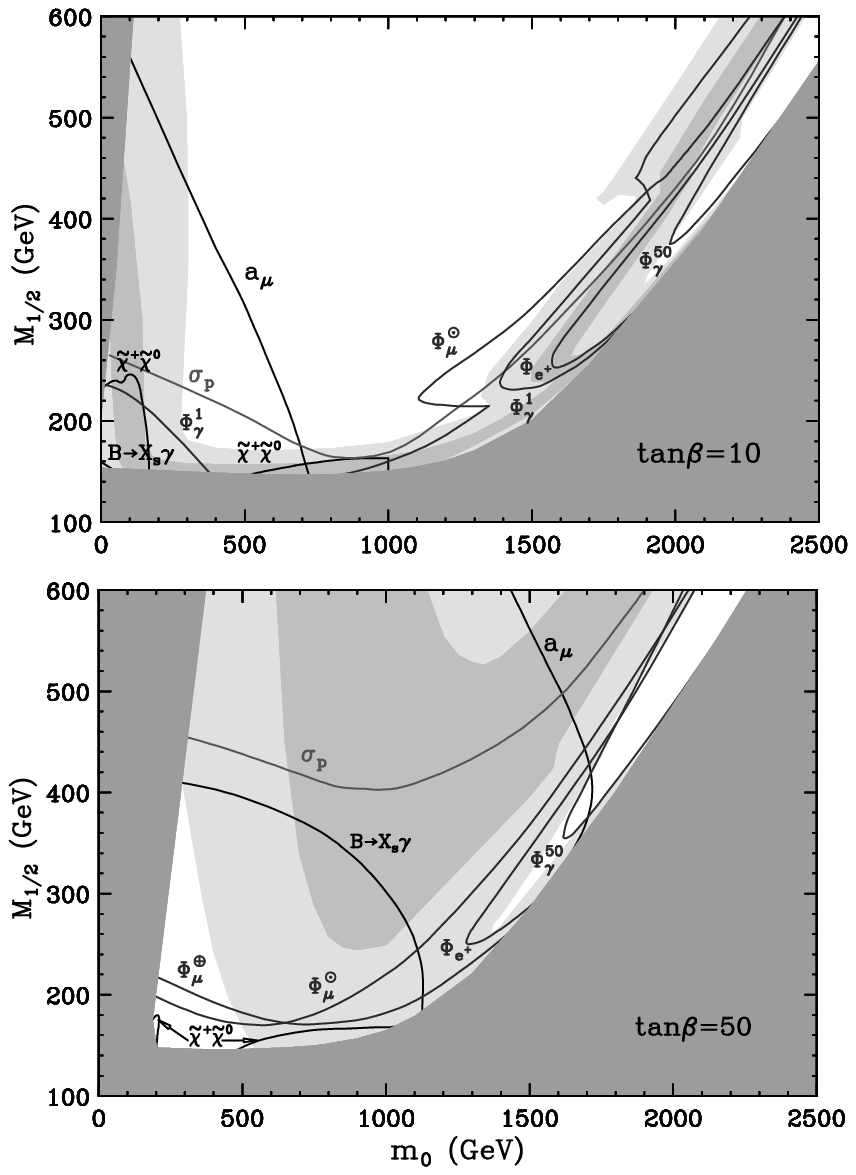}
\end{center}
\caption[]{SUSY parameter space explored by passive experiments.
Before the advent of LHC most of the cosmologically interesting
parameter space will be explored by ground based or space born
experiment. Fore more details see \cite{wilczek}} \label{wilcek}
\end{figure}

 In Figure \ref{electronpositron} we give an example  of   spectral distortions induced
 by $\chi$ annihilations on   the $e^+/(e^+  + e^-)$ ratio.  Quite interesting is the
 case of $\bar{D}$ production, since it has been suggested\cite{donato} that a $\bar{D}$  signal at
 kinetic energies  below $\sim 1\ GeV$ would be a strong indication for $\chi$ annihilation.
 In the case of high energy $\gamma$ rays the spectral deformation due to $\chi$ annihilation
 is expected to have a strong spatial dependence, mimicking the DM halo structure which might
 have more than one clumps in our galaxy. For a subset of SUSY models   this
 flux of HE $\gamma$ rays could be detectable by space born experiments (Figure \ref{wilcekethr}).

 For many SUSY models the indirect search for $\chi$  in CR will be a difficult undertaking:
 SUSY predictions depends in fact from several parameters and the limited precision in the knowledge of CR
 composition and spectra  will challenge the unambiguous detection of a  $\chi$ signal.
Before the advent of the LHC, however, the physics potential of
SUSY searches in space is promising\cite{wilczek,ellisneutralino}.
In fact most of the cosmologically relevant SUSY parameter space
can be explored by CR experiments. Capability to measure at the
same time all type of charged and neutral  particles which can be
signature of $\chi$ annihilation would clearly be an experimental
advantage: starting from 2006, the AMS-02 experiment on the ISS
 will be the only experiment able to precisely measure at the same time all kind of particles related to
$\chi$ decay. Other experiment will also  contribute to this
search: PAMELA will contribute by precisely measure the $\bar p$
component,  in particular at low energy, starting in 2003, while
GLAST  will perform the most precise measurement of the high
energy $\gamma$ ray halo structure and spectra starting in 2007.

\subsection{Indirect search for ultraheavy particles}

The existence of a measurable  flux  of Extreme Energies Cosmic
Rays  above $10^{19} eV$, represents a big puzzle in modern Astro
Particle physics. The behavior of the GKS cutoff versus energy and
particle type  shown in Figure \ref{ParticlePropagation} suggests
that no particle except the $\nu$'s can travel for large distances
at these energies. Above $10^{19} eV$ the volume of the region
which can be traversed by hadronic EECR is then dramatically
reduced.

One would then expect a steeper spectrum above the GZK cutoff. The
experimental results suggest instead a smoother spectrum which
would not be possible to explain using conventional physics.
Although we are talking only about few dozen events collected by
various experiments over several years, their existence challenge
standard explanations. One possible scenario is based on EE $\nu$
emitted in the decay of extremely heavy particles present in the
very early phases of the universe. These particle travel until
they reach regions close to us where they do interact with
ordinary IS matter and produce EECR which in turn reach our planet
where they can be detected. During this decade the advent of space
experiments like EUSO\cite{scarsi} and KLYPVE\cite{khrenov} will
improve by one or two orders of magnitude the sensitivity to EECR
of the Auger experiment, hopefully clarifying the present
situation.

\section{Conclusion}
One hundred years after their discovery  Cosmic Rays have still an
important potential for   new physics.  In order to exploit this
potential new, more sensitive experiments are planned which can
take advantage of the  unique conditions of space for precision
measurement of the primary CR flux. During the current decade
these experiments might well deliver exciting surprises in Astro
Particle physics, on issues like antimatter, dark matter or other
exotic states of matter.

\section{Acknowledgments}
 This work has been partially supported by the Italian Space Agency (ASI) under contract
 ARS-98/47.

%INDEX%%%%%%%%%%%%%%%%%%%%%%%%%%%%%%%%%%%%%%%%%%%%%%%%%%%%%%%%%%%%%%%
% Please check with the editor of your book whether he plans to
% include a "mutual" subject index - if so, please code your entries
% in the standard syntax. For your own purposes you may print your
% "personal" index by using the following commands:
%
%\clearpage
%\addcontentsline{toc}{section}{Index}
%\flushbottom
%\printindex
%%%%%%%%%%%%%%%%%%%%%%%%%%%%%%%%%%%%%%%%%%%%%%%%%%%%%%%%%%%%%%%%%%%%%

\end{document}